\documentclass[5p,twocolumn]{elsarticle}

\usepackage{graphicx,epsfig}
\usepackage{amsmath,mathrsfs,amsfonts}
\usepackage {amssymb}
\usepackage {longtable}
\usepackage{multirow}
 \usepackage{enumerate}

\usepackage{bm}
\usepackage{amsfonts}
\usepackage{subfigure}
\usepackage{color}
\usepackage{relsize}
 \usepackage[utf8]{inputenc}
\usepackage{hyperref}

\newcommand{\be}{\begin{equation}}
\newcommand{\ee}{\end{equation}}
\newcommand{\bea}{\begin{eqnarray}}
\newcommand{\eea}{\end{eqnarray}}

\newcommand{\abs}[1]{\lvert#1\rvert}

\journal{Journal of \LaTeX\ Templates}

\biboptions{sort&compress}
\begin{document}

\begin{frontmatter}

\title{Generating primordial fluctuations from modified teleparallel gravity with local Lorentz-symmetry breaking}


\author{Manuel Gonzalez-Espinoza\fnref{myfootnote}}
\author{Giovanni Otalora\fnref{myfootnote2}}
\address{Instituto de F\'{\i}sica, Pontificia Universidad Cat\'olica de 
Valpara\'{\i}so, 
Casilla 4950, Valpara\'{\i}so, Chile}
\fntext[myfootnote]{manuel.gonzalez@pucv.cl}
\fntext[myfootnote2]{giovanni.otalora@pucv.cl}


\begin{abstract}
In the context of modified teleparallel gravity, we study the generation of primordial density fluctuations in a general scalar-torsion theory whose Lagrangian density is an arbitrary function $f(T,\phi)$ of the torsion scalar $T$ and a scalar field $\phi$, plus the kinetic term of this latter. It is well known that generic modifications of teleparallel gravity are not invariant under six-parameter local Lorentz transformations. In order to restore the local Lorentz symmetry, we have incorporated six additional degrees of freedom in the form of Goldstone modes of the symmetry breaking through a Lorentz rotation of the tetrad field. After integrating out all the auxiliary modes, we obtain a second order action for the scalar and tensor propagating modes and their power spectrum generated during inflation.
It is found that an explicit mass term emerges in the second order action for curvature perturbation, describing the imprints of local Lorentz violation at first-order of slow-roll. We show that only inflationary models with nonminimal coupling functions $f(T,\phi)$ which are non-linear in $T$, including the case of $f(T)$ gravity with minimally coupled scalar field, can generate primordial fluctuations. For a concrete model of inflation, we study the power-law potential by using the latest Planck data.
\end{abstract}

%\begin{keyword}
%\texttt{elsarticle.cls}\sep \LaTeX\sep Elsevier \sep template
%\MSC[2010] 00-01\sep  99-00
%\end{keyword}

\end{frontmatter}

%\linenumbers

\section{Introduction}\label{Introduction}
%%%%%%%%%%%%%%%%%%%%%%%%%%%%%%%%%%%%%%%%%%%%%%%%%%%%%%%%%%%%%%%%%%%%%
Lorentz invariance is considered as one of the most fundamental symmetries in physics, which provides a fundamental support for the principles of general relativity (GR) and the standard model of particle physics \cite{Weinberg:1972kfs}. Nevertheless, at a sufficiently high-energy scale (Planck scale), it is expected that these two field theories merge into a single unified and quantum-consistent theory, under a possible breaking of local Lorentz symmetry \cite{Gasperini:1985aw,Kostelecky:1989jw,Colladay:1996iz,Kostelecky:2003fs,Bluhm:2004ep,Bluhm:2007bd,Bluhm:2017pje,Mattingly:2005re}. Furthermore, if primordial density fluctuations were generated during inflation \cite{Guth:1980zm,Starobinsky,Linde:1981mu,Mukhanov:1990me}, they give us an unique opportunity to learn about physics at energy scales that otherwise would not be accessible, since it is believed that inflation occurs near the scale of grand unification, and therefore, it is not too far from scales where quantum gravity is relevant \cite{MukhanovBook,Ackerman:2007nb,Baumann:2014nda}.

It is well known that modified gravity theories constructed from the so-called teleparallel equivalent of general relativity, or simply, teleparallel gravity (TG) \cite{Einstein,TranslationEinstein,Early-papers1,Early-papers2,Early-papers3,Early-papers4,Early-papers5,Early-papers6,JGPereira2,AndradeGuillenPereira-00,Arcos:2005ec,Pereira:2019woq}, break the local Lorentz symmetry \cite{Sotiriou:2010mv,Li:2010cg}. This has caused great interest in the study of cosmic inflation and the effects of local Lorentz violation on the inflationary observables from the framework of modified teleparallel gravity (MTG) theories \cite{Ferraro:2006jd,Wu:2011kh,Rezazadeh:2015dza,Wu:2016dkt,Gonzalez-Espinoza:2019ajd,Raatikainen:2019qey}. Particularly, in \cite{Rezazadeh:2015dza} the authors have investigated power-law and intermediate inflation in $f(T)$ gravity, whereas in \cite{Gonzalez-Espinoza:2019ajd} it has been studied the slow-roll inflation in a generalized non-minimally coupled scalar-torsion gravity theory with a Galileon-type self-interaction. In Ref. \cite{Wu:2016dkt}, the authors concentrated their efforts in the investigation of the consequences of local Lorenz violation to the generation of primordial density fluctuations. They showed that due to local Lorentz violation no subhorizon scalar-perturbation mode can survive by the time of horizon crossing, and thus these theories are incapable of generating enough primordial density inhomogeneity, even if it brings some de Sitter background solution. Moreover, in \cite{Raatikainen:2019qey}, where a nonminimal coupling to the vector torsion has been included, the authors have corroborated the aforementioned result of \cite{Wu:2016dkt}, and they have also concluded that for some relation between the coupling functions to torsion scalar and vector torsion, the scalar field can source the linear perturbations.

We study the generation of primordial density fluctuations in a larger class of generalized teleparallel scalar-torsion  $f(T,\phi)$ gravity theories with local Lorentz symmetry breaking. We confirm what was obtained in \cite{Wu:2016dkt} in regards the case of a nonminimal coupling function $f(T,\phi)$ lineal in $T$, but we also show that inflationary models with nonminimal coupling function $f(T,\phi)$ which is non-linear in $T$, can generate primordial fluctuations. The paper is organized as follows. In section \ref{TG}, we give a concise introduction to TG. In section \ref{GS_torsion_gravity}, we develop the framework of generalized $f(T,\phi)$ gravity theory, calculating the background equations for a Friedmann-Robertson-Walker (FRW) metric, and then analysing the de-Sitter limit. In the sections \ref{scalar_pert} and \ref{tensor_pert} , we investigate the cosmological perturbations using the ADM formalism for the tetrad fields and using the Maldacena's method of expanding the action until second order for the perturbations \cite{Maldacena:2002vr}. In section \ref{Chaotic}, we apply our results to the particular case of the power-law potential. Finally, in Section \ref{Concluding_Remarks}, we summarize our findings and present our main conclusions and final remarks.

%%%%%%%%%%%%%%%%%%%%%%%%%%%%%%%%%%%%%%%%%%%%%%%%%%%%%%%%%%%%%%%%%%%%

\section{Teleparallel Gravity}\label{TG}
%%%%%%%%%%%%%%%%%%%%%%%%%%%%%%%%%%%%%%%%%%%%%%%%%%%%%%%%%%%%%%%%%%%%

Teleparallel Gravity (TG) is a gauge theory for the translation group which constitutes an alternative description of gravity based on torsion \cite{JGPereira2,AndradeGuillenPereira-00,Arcos:2005ec,Pereira:2019woq}. The dynamical variable of TG is the tetrad field $\mathbf{e}^{}_{A}(x^\mu)$, and it connects the spacetime metric $g_{\mu\nu}$ and the Minkowski tangent space metric $\eta _{AB}^{}=\text{diag}\,(-1,1,1,1)$ thorough the local relation 
\begin{equation}  \label{metrics}     % {1} ==>
 g_{\mu\nu}=e^{A}_{~\mu}\,e^{B}_{~\nu}\,\eta_{AB}^{}\,,
 \end{equation} where $e^{A}_{~\mu}$ are the tetrad components in a coordinate base and also satisfying the orthogonality conditions  $e^{A}_{~\mu} e_{A}^{~\nu}=\delta^{\nu}_{\mu}$ and $e^{A}_{~\mu} e_{B}^{~\mu}=\delta^{A}_{B}$, with $e_{B}^{~\mu}$ the inverse components.

The action functional of TG is given by
\be
S=-\frac{M_{pl}^2}{2}\int{d^{4}x\,e\,{T}},
\label{action_TG}
\ee being $T$ the torsion scalar, $e=\det{\left(e^{A}_{~\mu}\right)}=\sqrt{-g}$, and $M_{pl}^2=\left(8\pi G\right)^{-1}$ the reduced Planck mass. The torsion scalar is defined as 
\be
T= S_{\rho}^{~\mu\nu}\,T^{\rho}_{~\mu\nu},
\label{ScalarT}
 \ee
 where 
\begin{equation} \label{Def_Torsion}   %% MR  \label{4}
 T^{\rho}_{~\mu\nu}\equiv e_{A}^{~\rho}\left[\partial_{\mu}e^{A}_{~\nu}
 -\partial_{\nu}e^{A}_{~\mu}+\omega^{A}_{~B\mu}\,e^{B}_{~\nu}
 -\omega^{A}_{~B\nu}\,e^{B}_{~\mu}\right]\,,
\end{equation} are the components of torsion tensor, and
\begin{equation} \label{Superpotential}
 S_{\rho}^{~\mu\nu}=\frac{1}{2}\left(K^{\mu\nu}_{~~\rho}+\delta^{\mu}_{~\rho} \,T^{\theta\nu}_{~~\theta}-\delta^{\nu}_{~\rho}\,T^{\theta\mu}_{~~\theta}\right)\,,
\end{equation} is the so-called super-potential, with
\begin{equation}  \label{Contortion}
 K^{\mu\nu}_{~~\rho}= -\frac{1}{2}\left(T^{\mu\nu}_{~~\rho}
 -T^{\nu\mu}_{~~\rho}-T_{\rho}^{~\mu\nu}\right).
\end{equation} the contorsion tensor.  

The spin connection of TG is written as 
\be
\omega^{A}_{~B \mu}=\Lambda^{A}_{~D}(x) \partial_{\mu}{\Lambda_{B}^{~D}(x)},
\label{spin_TG}
\ee with $\Lambda^{A}_{~D}(x)$ a local (point-dependent) Lorentz transformation \cite{Aldrovandi-Pereira-book}. Let us remember that Eq. \eqref{metrics} only determines locally the tetrad frame up to transformations of six-parameter Lorentz group in the tangent spaces indices \cite{JGPereira2,AndradeGuillenPereira-00,Arcos:2005ec}. In this way, there are infinity of these six-fold tetrads, each one relating the Minkowski tangent space metric $\eta_{AB}$ to the same spacetime metric $g_{\mu\nu}$. Thus, as in the tetrad formalism for GR, we need to introduce a Lorentz connection in order to ensure the covariance of the theory at hand \cite{Weinberg:1972kfs}. However, unlike the spin connection of GR, which represents both gravitational and inertial effects, the spin connection of TG as defined in \eqref{spin_TG}, represents only inertial effects of the frame, that is to say, it is a flat spin connection \cite{Aldrovandi-Pereira-book}. So, there is a special class of inertial frames, called the proper frames in which it vanishes, $\omega^{A}_{~B \mu}=0$. Therefore, starting from this class of inertial frames we can obtain all the other classes of frames by performing local (point-dependent) Lorentz transformations on both, the tetrad and the spin connection \cite{Aldrovandi-Pereira-book}. When solving the gravitational field equations for the tetrad field, the local Lorentz invariance of TG allows us to choice a particular class of frames, i.e. the preferred class of inertial frames $\omega^{A}_{~B \mu}=0$ for the sake of simplicity. Equivalently, it can be seen that the spin connection enters into the action of TG through a total derivative term and, therefore, the solution to the gravitational field equations does not depend on the spin connection \cite{Krssak:2018ywd}

The corresponding spacetime-indexed linear connection is 
\be
\Gamma^{\rho}_{~\mu\nu}=e_{A}^{~\rho}\left(\partial_{\nu}{e^{A}_{~\mu}}+\omega^{A}_{~B \nu}e^{B}_{~\mu}\right),
\ee
 which is the so-called Weitzenb\"{o}ck connection. It is related to the Levi-Civita connection of GR through
\be
\Gamma^{\rho}_{~\mu\nu}=\bar{\Gamma}^{\rho}_{~\mu\nu}+K^{\rho}_{~\mu\nu}.
\label{Gamma_relation}
\ee Using this latter equation, it can be shown that 
\be
T=-R-e^{-1} \partial_{\mu}{\left(e T^{\nu \mu}_{~~~\nu}~\right)},
\ee  where $R$ is the curvature scalar of Levi-Civita connection \cite{Aldrovandi-Pereira-book}. This equation show that TG and GR are equivalent theories in the level of field equations. 

However, when one modifies gravity from the viewpoint of TG, by introducing a non-minimally coupled matter field, as for example a scalar field \cite{Geng:2011aj,Otalora:2013tba,Otalora:2013dsa,Otalora:2014aoa,Skugoreva:2014ena}, or by adding into the action, non-linear terms in the torsion scalar $T$, as for example in $f(T)$ gravity \cite{Bengochea:2008gz,Linder:2010py,Li:2011wu,Gonzalez-Espinoza:2018gyl}, it is obtained a new class of modified gravity theories with a rich phenomenology and not equivalent to their corresponding counterpart based on curvature \cite{Cai:2015emx}.

Below, we are going to study the generation of primordial density fluctuations in generalized teleparallel scalar-torsion gravity theories. 

\section{Generalized Scalar-torsion gravity}\label{GS_torsion_gravity}

\subsection{Field equations and local Lorentz invariance}

The relevant action is given by 
\begin{equation}
 S=\int d^{4}x\,e\,\left[ f(T,\phi)+ P(\phi)X \right],
\label{action1}
\end{equation}
where  $f$ is an arbitrary function of $\phi$ and $T$, and also $X=-\partial_{\mu}{\phi}\partial^{\mu}{\phi}/2$. This general action includes non-minimally coupled scalar-torsion gravity models with $f(T,\phi)$ the coupling function, and $f(T)$ gravity, plus minimally coupled scalar field. For $f(T,\phi)=-M_{pl}^2 T/2-V(\phi)$, we recover TG, with $V(\phi)$ the scalar potential \cite{MukhanovBook}.

Varying the action with respect to the tetrad field $e^{A}_{~\mu}$ we find the corresponding field equations
\bea
&&f_{,T} G_{\mu \nu}+S_{\mu \nu}{}^{\rho} \partial_{\rho} f_{,T}+\frac{1}{4}\delta_{\mu \nu}\left(f-T f_{,T}\right)+\nonumber\\
&& \frac{P}{4}\left(\delta_{\mu \nu} X+\partial_{\mu}\phi \partial_{\nu}\phi\right)=0,
\label{FieldEquations}
\eea which have been expressed in a general coordinate basis, and $G^{\mu}_{~\nu}=e_{A}^{~\mu} G^{A}_{~\nu}$ is the Einstein tensor, and 
the tensor $G_{A}^{~\mu}$ is defined as $G_{A}^{~\mu}\equiv e^{-1}\partial_{\nu}\left(e e_{A}^{~\sigma} S_{\sigma}^{~\mu\nu}\right)-e_{A}^{~\sigma} T^{\lambda}_{~\rho \sigma}S_{\lambda}^{~\rho \mu}+e_{B}^{~\lambda} S_{\lambda}^{~\rho \mu}\omega^{B}_{~A \rho}+\frac{1}{4}e_{A}^{~\mu} T$ \cite{Aldrovandi-Pereira-book}. The equation \eqref{FieldEquations} has an antisymmetric part associated with the tensor $S_{\mu \nu}{}^{\rho}$ in the second term. It is an expected result as the action \eqref{action1} is not local Lorentz invariant \cite{Sotiriou:2010mv,Li:2010cg}. To see this explicitly, let us consider the infinitesimal point-dependent Lorentz transformation $e'^{A}_{~\mu}=e^{A}_{~\mu}+\xi_{B}^{~A}e^{B}_{~\mu}$, with $\xi^{A B}=-\xi^{B A}$. Under this transformation the variation of the action is 
\be
\delta{S}=\int{d^{4}{x}e \partial_{\rho} f_{,T} S_{\mu \nu}{}^{\rho} \xi^{\mu\nu}},
\ee where $\xi^{\mu\nu}=e_{A}^{~\mu} e_{B}^{~\nu}\xi^{A B}$. The condition $\delta{S}=0$ for arbitrary $\xi^{A B}$ leads us to the following constraint
\be
\partial_{\rho} f_{,T} S_{\left[\mu \nu\right]}{}^{\rho}=0.
\label{Constr}
\ee For TG, $f\sim T$, one has $\partial_{\rho} f_{,T}=0$, and thus the left hand side of this latter equation becomes identically equal to zero, and local Lorentz invariance is restored \cite{Aldrovandi-Pereira-book}. For MTG, $\partial_{\rho} f_{,T} \neq 0$, it corresponds to a set of six equations for six additional degrees of freedom, due to violation of local Lorentz symmetry.

\subsection{Cosmological background}

We impose the standard homogeneous and isotropic background geometry by choosing
\begin{equation}
\label{veirbFRW}
e^A_{~\mu}={\rm
diag}(1,a,a,a),
\end{equation}
which corresponds to a flat Friedmann-Robertson-Walker
(FRW) universe with metric 
\begin{equation}
ds^2=-dt^2+a^2\,\delta_{ij} dx^i dx^j \,,
\label{FRWMetric}
\end{equation}
where $a$ is the scale factor which is a function of the cosmic time $t$. In MTG, due the violation of local Lorentz invariance \cite{Sotiriou:2010mv,Li:2010cg}, the gravitational field equations and their solution become dependent on the spin connection and, therefore, it is necessary to know a way to retrieve the corresponding spin connection associated with each tetrad field in order to correctly solve the field equations. Particularly, in the context of FRW cosmologies it has been shown that diagonal tetrad \eqref{veirbFRW} is a proper tetrad, that is to say, the appropriated spin connection associated with this tetrad is the vanishing spin connection, which leads to results of physical meaning \cite{Krssak:2015oua}.

Replacing the tetrad field \eqref{veirbFRW} for the vanishing spin connection into the field equations \eqref{FieldEquations}, we obtain the background equations
\bea
\label{00}
 f(T,\phi) - P(\phi) X - 2 T f_{,T}=0, \\
\label{ii}
 f(T,\phi) + P(\phi) X - 2 T f_{,T} - 4 \dot{H} f_{,T} - 4 H \dot{f}_{,T}=0, \\
\label{phi}
 - P_{,\phi} X - 3 P(\phi) H \dot{\phi} - P(\phi) \ddot{\phi} + f_{,\phi}=0,
\eea
where $H\equiv \dot{a}/a$ is the Hubble rate, and a dot represents derivative with respect to $t$. Also, a comma denotes derivative with respect to $\phi$ or $T$. In these equations, we also have $T=6 H^2$, which has been obtained from Eq. \eqref{ScalarT}. 

To analyse the de-Sitter limit \cite{Jarv:2015odu}, we use the values $\phi=\phi_{*}$ and $H=H_{*}(t)$ in the background equations, so we obtain $\dot{H}_{*} = 0$ and $f_{,\phi}(T_{*},\phi_{*}) = 0$. After that, we study the perturbations of this de-Sitter limit, using $\phi=\phi_{*}+\delta \phi(t)$ and $H=H_{*}+\delta H(t)$. Then, we expand equation \eqref{00} to first order and we find $\delta H = 0$. Next, expanding equation \eqref{phi} to first order, we find the equation for $\delta \phi(t)$,
\begin{equation}
\delta{\ddot{\phi}} + 3 H_{*} \delta{\dot{\phi}} - \dfrac{f_{,\phi\phi}(T_{*},\phi_{*})}{P(\phi_{*})} \delta \phi = 0,
\end{equation}
whose solution is given by,
\begin{equation}
\delta \phi(t) =C_{1} e^{\mu_{+} t}+C_{2} e^{\mu_{-} t},
\end{equation}
where $C_{1}$ and $C_{2}$ are integration constants, and
\be
\mu_{\pm}=\frac{-3 H_{*}}{2}\left[1\pm \sqrt{1+\frac{4 f_{,\phi\phi}(T_{*},\phi_{*})}{9 H_{*}^{2} P(\phi_{*})}}\right].
\ee

Therefore, the perturbation $\delta \phi$ is stable only for
\begin{equation}
\dfrac{f_{,\phi\phi}(T_{*},\phi_{*})}{P(\phi_{*})} < 0.
\label{deSitterStability}
\end{equation} For non-phantom scalar fields, it is required $P(\phi_{*})>0$, and then the above constraint becomes $f_{,\phi\phi}(T_{*},\phi_{*})<0$. For $f(T)$ gravity, without dynamical scalar field, one has $f_{,\phi\phi}(T_{*},\phi_{*})=0$, and then the eigenvalues are $\mu_{-}=0$ and $\mu_{+}=-3 H_{*}$. Therefore, in this latter particular case, the de-Sitter background is always (marginally) stable \cite{Cai:2015emx}.

In order to realize the slow-roll approximation into the present scenario, from Eqs. \eqref{00} and \eqref{ii} we obtain
\begin{equation}
\epsilon =  \delta_{PX} + \delta_{f_{,T}},
\label{slow_roll_eps}
\end{equation}
where we have introduced the slow-roll parameters
\bea
&& \epsilon= -\dfrac{\dot{H}}{H^2},\:\:\: \delta_{P X} = - \dfrac{P(\phi) X}{2 H^{2} f_{,T}},\:\:\ \delta_{f_{,T}} = \dfrac{\dot{f}_{,T}}{f_{,T} H}.
\label{slowpara1}
\eea Also, it is useful to split the parameter $\delta_{f_{,T}}$ as
\be
\delta_{f_{,T}}=\delta_{f\dot{H}}+\delta_{fX},
\label{deltafT}
\ee where we define
\be
\delta_{f\dot{H}}=\frac{f_{,TT} \dot{T}}{H f_{,T}},\:\:\:\: \delta_{fX}=\frac{f_{,T\phi}\dot{\phi}}{H f_{,T}}.
\ee
Thus, from the relations \eqref{slow_roll_eps} and \eqref{slowpara1}, it is easy to obtain 
\be
\delta_{f\dot{H}}=-\frac{2 \mu}{1+2\mu}\left(\delta_{PX}+\delta_{fX}\right), 
\ee
\be
\delta_{f_{,T}}=\frac{1}{1+2\mu} \left(\delta_{fX}-2 \mu \delta_{PX}\right)
\ee and then
\be
\epsilon=\frac{1}{1+2\mu}\left(\delta_{PX}+\delta_{fX}\right),
\ee where we have defined $\mu\equiv T f_{,TT}/f_{,T}$ in analogy with the deviation parameter of the (curvature based) modified gravity theories \cite{DeFelice:2010aj}.

The time dependence of the slow-roll parameters is calculated as 
\bea
&& \frac{\dot{\delta}_{PX}}{H \delta_{PX}}=\delta_{P}+2 \delta_{\phi}+2 \epsilon-\delta_{f_{,T}},\\
&& \frac{\dot{\delta}_{f_{,T}}}{H \delta_{f_{,T}}}=\delta_{f_{,TT}}+\delta_{PX}+\left(\delta_{f_{,T\phi}}-\delta_{f_{,TT}}\right) \frac{\delta_{fX}}{\delta_{f_{,T}}},\\
&& \frac{\dot{\delta}_{fX}}{H \delta_{fX}}=\delta_{f_{,T\phi}}+\delta_{\phi}+\delta_{PX},
\eea  where it has also been defined the slow-roll parameters
\bea
&& \delta_{P}=\frac{\dot{P}}{H P},\:\:\:\: \delta_{\phi}=\frac{\ddot{\phi}}{H\dot{\phi}},\nonumber\\
&& \delta_{f_{,TT}}=\frac{\dot{f}_{,TT}}{H f_{,TT}},\:\:\:\delta_{f_{,T\phi}}=\frac{\dot{f}_{,T\phi}}{H f_{,T\phi}}. \label{slowpara_n1}
\eea During slow-roll inflation $\dot{\delta}_{PX}\sim \dot{\delta}_{f,T}\sim \dot{\delta}_{fX}\sim \mathcal{O}(\epsilon^2)$, and similarly for the other parameters. Furthermore, using these slow-roll parameters we can write $\dot{\mu}/(\mu H) = - 2\epsilon + \delta_{f,_{TT}} - \delta_{f,_{T}}$. Consequently, the fractional change of $\mu$ per Hubble time is small, i.e., $\left|\dot{\mu}/(H \mu)\right| \ll 1$, and then $\mu$ is nearly constant during slow-roll inflation \cite{Baumann:2009ds}. It is important to notice that a substantial fractional change of $\mu$ per Hubble time would indicate the failure of the slow-roll approximation.

\section{Scalar Perturbations}\label{scalar_pert}

\subsection{Second order action}

In order to study primordial density fluctuations, we start from the Arnowitt-Deser-Misner (ADM) decomposition of the tetrad field \cite{Wu:2011kh}
\bea
&& e^{0}_{~\mu}=\left(N,\textbf{0}\right),\:\:\:\: e^{a}_{~\mu}=\left(N^{a},h^{a}_{~i}\right)\label{ADM1},\\
&& e_{0}^{~\mu}=\left(1/N,-N^{i}/N\right),\:\:\:\: e_{a}^{~\mu}=\left(0, h_{a}^{~i}\right)\label{ADM2},
\eea where $N^{i}=h_{a}^{~i}  N^{a}$, with $h^{a}_{~j} h_{a}^{~i}=\delta^{i}_{j}$, being $h^{a}_{~i}$ the induced tetrad field.

Using the uniform field gauge, $\delta \phi=0$, a convenient ansatz for the fields is 
\be
N=1+\alpha,\:\:\:\: N^{a}=a^{-1} e^{-\mathcal{R}}\delta^{a}_{~i} \partial^{i}{\psi},\:\:\:\: h^{a}_{~i}=a e^{\mathcal{R}}\delta^{a}_{~j}\delta^{j}_{~i},
\label{Uniform_Field_Gauge}
\ee which gives the corresponding perturbed metric \cite{DeFelice:2011uc}
\bea
 ds^2 &=&-\left[\left(1+\alpha\right)^2-a^{-2}e^{-2\mathcal{R}}\left(\partial \psi\right)^2\right]dt^2 \nonumber \\ 
&& +2\partial_{i}{\psi}dt dx^{i}+ a^2 e^{2\mathcal{R}}\delta_{i j} dx^{i} dx^{j}.
\eea

At the perturbation level, a suitable procedure (widely used in the literature) to handle the violation of local Lorentz invariance in MTG theories, consists in adding the corresponding six Lorentz degrees of freedom, related to the breaking of local Lorentz symmetry, directly into the perturbed tetrad field \eqref{Uniform_Field_Gauge} \cite{Izumi:2012qj,Golovnev:2018wbh}. These additional modes are gauge degrees of freedom in the Lorentz-invariant limit and contain the dependence with the inertial effects of the frame when choosing the particular perturbed tetrad frame \eqref{Uniform_Field_Gauge}, among the infinite choices, and thus leading us to results of physical meaning at the moment of solving the perturbed field equations \cite{Bluhm:2004ep}. Therefore, they can be incorporated in the form of Goldstone modes of the symmetry breaking, by performing a Lorentz rotation of the tetrad field \cite{Bluhm:2004ep, Bluhm:2007bd}.  So, under the transformation 
\be
\Lambda^{A}_{~B}=\left(e^{\chi}\right)^{A}_{~B}=\delta^{A}_{~B}+\chi^{A}_{~B}+\frac{1}{2} \chi^{A}_{~C} \chi^{C}_{~B}+\mathcal{O}(\chi^3),
\label{Lorentz_Transf}
\ee and keeping fixed the zero spin connection for the cosmological background, the full tetrad field is written as 
\bea
 e'^{A}_{~\mu}&=&\left(e^{\chi}\right)^{A}_{~B} e^{B}_{~\mu},\nonumber\\
 &=&e^{A}_{~\mu}+\chi^{A}_{~B}e^{B}_{~\mu}+\frac{1}{2} \chi^{A}_{~C} \chi^{C}_{~B} e^{B}_{~\mu}+\mathcal{O}(\chi^3).
\label{Transf_tetrad}
\eea
The matrix $\chi_{A B}=-\chi_{B A}$ is parametrized as 
\be
\chi^{0}_{~B}=\left(0,\chi_{b}\right),\:\:\:\: \chi^{a}_{~B}=\left(\chi^{a}, B^{a}_{~b}\right),
\ee where $\chi^{a}=\eta^{a b}\chi_{b}$ and  $B_{ab}=-B_{ba}$. It is defined the spatial vector $\chi^{i}=h_{a}^{~i} \chi^{a}=\partial_{i}{\beta}+\chi^{(T)}_{i}$, and the spatial antisymmetric tensor $B_{i j}=h^{a}_{~i} h^{b}_{~j} B_{a b}=-B_{j i}=-\epsilon_{j i k} B^{k}$. Therefore, there are a scalar mode $\beta$, a transverse vector mode $\chi^{(T)}_{i}$ and a (pseudo) vector mode $B_{i}$ \cite{Wu:2016dkt,Golovnev:2018wbh}.

Here it may be important to make some clarifications. It has been shown in Ref. \cite{Krssak:2015rqa} that the purely inertial spin connection of TG, Eq. \eqref{spin_TG}, has the function of removing the fake contributions coming from inertial effects contained in the gravitational action. The authors have figure out a method to find, for each tetrad, a naturally associated spin connection which locally removes the inertial effects of the action and thus providing a renormalization process for TG. This method also has been extended in Ref. \cite{Krssak:2015oua} to be applied in the context of MTG where it has been used to construct a covariant formulation of $f(T)$ gravity. The key difference between the method used in TG and that implemented in covariant $f(T)$ gravity to determine the spin connection, lies in the fact that for TG we can first solve the gravitational field equations for a particular class of frames (i.e. the inertial class for the vanishing spin connection) in order to find a reference tetrad where gravity is switched-off, and then to get the appropriated spin connection, while in the case of MTG, we can only guess the reference tetrad from arguments of symmetry based on the knowledge of the coordinate system in which the ansatz tetrad is written \cite{Krssak:2018ywd}.

Thus, when studying cosmological perturbations from this covariant formalism for MTG, it can be expected that the spin connection continues playing the role of representing the inertial effects of the perturbed tetrad frame, as it is the foundation of its definition in the context of TG \cite{Aldrovandi-Pereira-book}. However, the retrieve of the appropriated spin connection for the perturbed tetrad frame \eqref{Uniform_Field_Gauge}, using the method established in \cite{Krssak:2015oua}, is not apparent, and as far as we know, it has not been done in the literature. On the other hand, one might think to perturb the spin connection in addition to the tetrad field. In this case the perturbed spin connection should carry out the six additional Lorentz degrees of freedom. For example, we can use the transformation \eqref{Lorentz_Transf} into \eqref{spin_TG} in order to find the perturbed spin connection around the vanishing spin connection. Nevertheless, this is equivalent to directly adding the Lorentz degrees of freedom to the perturbed tetrad \eqref{Uniform_Field_Gauge}, while keeping the vanishing spin connection of the background without a perturbation, such as we have done in Eq. \eqref{Transf_tetrad} \cite{Golovnev:2020aon}. This is because a same Lorentz rotation in both, the tetrad and the spin connection, leaves invariant the torsion tensor as defined in Eq. \eqref{Def_Torsion} \cite{Aldrovandi-Pereira-book}.

Following \cite{Maldacena:2002vr}, the next step is to expand the action \eqref{action1} up to second order to obtain 
\begin{equation}
\begin{array}{lll}
S^{(2)}&=& \mathlarger{\int} dt d^{3}x a^3\left[ \dfrac{2}{a^2} ( w_1 \dot{\mathcal{R}} - w_1 H \alpha ) \partial^2 \psi + 6 w_1 H \alpha \dot{\mathcal{R}}\right. 
\\
&& -\dfrac{2 w_2 }{a^2} \alpha \partial^2 \mathcal{R} + w_3 \alpha^2 - 3 w_1 \dot{\mathcal{R}}^2 + \dfrac{w_2}{a^2}(\partial \mathcal{R})^2  \\
&& - 4 \left(   w_4 \dot{\mathcal{R}}  - w_4 H \alpha \right) \partial^2 \beta +w_5 \mathcal{R} \partial^2 \beta +  w_{6} (\partial^2 \beta)^{2}   \\
&& \left. + \dfrac{w_{6}}{a^{4}} (\partial^2 \psi)^{2} - \dfrac{2 w_{6}}{a^2} (\partial^2 \beta \partial^2 \psi)  \right],  
\label{second_order}
\end{array}
\end{equation}
where we have defined the functions
\begin{eqnarray}
w_1&=&  -2 ( f_{,T} + 2 T f_{,TT}), \nonumber
\\
w_2&=& - 2 f_{,T}  ,\nonumber
\\
w_3&=& P(\phi)X + T f_{,T} + 2 T^2 f_{,TT},  \nonumber
\\
w_4 &=& -2 ( f_{,T} + T f_{,TT}), \nonumber
\\
w_5 &=& 4 \dot{f}_{,T}, \nonumber
\\
w_6 &=& \dfrac{4}{3} T f_{,TT} .
\end{eqnarray}
From action \eqref{second_order}, it can be seen that the scalar modes $\alpha$, $\psi$ and $\beta$ are auxiliary fields and do not propagate. Varying this action with respect to $\partial^{2}\psi$ leads us to
\begin{eqnarray}
w_{1} \dot{\mathcal{R}} - w_{1} H \alpha + \dfrac{w_{6}}{a^{2}} \partial^{2} \psi - w_{6} \partial^{2} \beta = 0,
\label{var1} 
\end{eqnarray}
whereas variation with respect to $\partial^{2} \beta$ gives
\begin{eqnarray}
&&-4 w_{4} \dot{\mathcal{R}} + 4 w_{4} H \alpha +  w_{5} \mathcal{R} -2 \dfrac{w_{6}}{a^{2}} \partial^{2} \psi\nonumber\\
&& + 2 w_{6} \partial^{2} \beta = 0, 
\label{var2}
\end{eqnarray}
and for $\alpha$ we have
\begin{eqnarray}
&&- 2 \dfrac{w_{1}}{a^{2}} H \partial^2 \psi + 6 w_{1} H \dot{\mathcal{R}} - 2 \dfrac{w_{2}}{a^{2}} \partial^2 \mathcal{R} + 2 w_{3} \alpha  \nonumber \\
&&+ 4 w_{4} H \partial^2 \beta = 0 \label{var3}.
\end{eqnarray}
Solving the above three equations for $\alpha$, $\partial^2{\psi}$ and  $\partial^2{\beta}$, and after substituting these results in Eq. \eqref{second_order}, the second order action for curvature fluctuation can be written as 
\begin{equation}
S^{(2)} = \int dt d^{3}x a^3 Q_{s} \left[ \dot{\mathcal{R}}^{2} - \dfrac{c^{2}_{s}}{a^{2}} (\partial \mathcal{R})^2 -m^{2} \mathcal{R}^2 \right],  
\label{SecOrderSM}
\end{equation}

where,
\begin{eqnarray}
\label{Qs}
Q_s &=& \dfrac{3 w_{1} H^{2} + w_{3}}{H^{2}}=\frac{PX}{H^2}, 
\\
\label{cs}
c^{2}_{s}&=& 1  ,
\\
m^{2}&=& \dfrac{\dot{w_{2}}}{w_{2}} \left(  3 H + \dfrac{\dot{Q}_{s}}{Q_{s}}  - 2 \dfrac{\dot{w_{2}}}{w_{2}} + \dfrac{\ddot{w_{2}}}{\dot{w}_{2}} + \dfrac{w_{1}\dot{w_{2}}}{w_{6} Q_{s}} \right) .  
\label{m2}
\end{eqnarray} 
The first and second term in action \eqref{SecOrderSM} are the usual terms appearing in the quadratic action of perturbations, while the third term is a new explicit mass term, that represents the effects of local Lorentz-symmetry breaking. The origin of this mass term is the Lorentz violating coupling term $f(T,\phi)$ in action \eqref{action1}. The emergence of this propagating massive scalar mode could be related to an alternative gravitational Higgs mechanism \cite{Kostelecky:1989jw,Bluhm:2007bd}.

For any theory to be physically viable, it must be free of ghosts and Laplacian instabilities by requiring $Q_{s}>0$ and $c^2_{s}>0$. These two conditions are satisfied by equations \eqref{Qs} (for $P>0$) and \eqref{cs}. Moreover, in the presence of an explicit mass term, there is an additional condition that is the non-occurrence of tachyonic instability \cite{DeFelice:2016ucp}. There are two situations in which the tachyonic instability can be avoided. The first possibility is that the mass squared must be positive, $m^2>0$, and, the second one, if we have $m^2<0$, then it is required that $\abs{m^2}\lesssim H^2$ \cite{DeFelice:2016ucp,Frusciante:2018vht}.

In terms of slow roll parameters we can write 
\begin{equation}
Q_{s} =w_{2} \delta_{PX},
\end{equation}
and it is also useful to define 
\be
\eta=\frac{\dot{Q}_{s}}{H Q_{s}}=\delta_{P}+2 \delta_{\phi}+2 \epsilon.
\ee 
Similarly, the mass term can be written as
\bea
&&\eta_{\mathcal{R}}=\frac{m^2}{3 H^2}=\delta_{f_{,T}}\left[1 + \left(1+\frac{\delta_{fX}}{\delta_{PX}}\right)\dfrac{\delta_{f_{,T}}}{\delta_{f\dot{H}}} \right],
\label{mass_term} 
\eea  For $f(T,\phi)$ non-linear in $T$, and either $\delta_{fX}= 0$ or $\delta_{fX}\neq 0$, one has that $\eta_{\mathcal{R}}\sim \mathcal{O}(\epsilon)$ is non-zero (and finite). Furthermore, tachyonic instability is avoided as long as $\abs{\eta_{\mathcal{R}}}\lesssim 1$.
In the absence of coupling between $T$ and $\phi$, one has $\delta_{fX}=0$, and thus $\delta_{f_{,T}}=\delta_{f\dot{H}}$. So, from Eq. \eqref{mass_term}, we find $\eta_{\mathcal{R}}=2 \delta_{f\dot{H}}\sim \mathcal{O}(\epsilon)$. This is the explicit mass term arising in $f(T)$ gravity, plus scalar field. For TG, $f\sim T$, one has $\delta_{f\dot{H}}=0$, and then $\eta_{\mathcal{R}}=0$, which is an expected result since TG is local Lorentz invariant \cite{Aldrovandi-Pereira-book}. 

Now, let us consider the case of $f(T,\phi)$ a linear function in $T$, and $\delta_{fX}\neq 0$. This is precisely the non-minimally coupled scalar-torsion theory of Ref. \cite{Wu:2016dkt}. For this model one has $\delta_{f\dot{H}}=0$, $\delta_{f_{,T}}=\delta_{fX}$, and then $\abs{\eta_{\mathcal{R}}}= \infty$. The physical meaning of this is that there are no nonzero-momentum solutions for the scalaron, as in this case one would have $\partial^{2}{\mathcal{R}}=0$, from action \eqref{second_order}, and then, spoiling the generation of primordial density fluctuations. This latter result is consistent with what was obtained in \cite{Wu:2016dkt}. 

\subsection{Mukhanov-Sasaki equation}
It is introduced the canonically-normalized Mukhanov variable
\begin{equation}
v \equiv z \mathcal{R},
\end{equation}
where we have also defined
\be
z^{2} =2 a^{2} Q_{s}. 
\ee 

Making the change to conformal time $d\tau=dt/a$, and using the above variables, action \eqref{SecOrderSM} can be written as

\begin{equation}
S^{(2)} =\frac{1}{2}\int d\tau d^{3}x \left[ (v')^{2} - c_{s}^{2} (\partial v)^{2} - M^{2}  v^{2} \right],
\label{SOA_v}
\end{equation} where it has been defined the effective mass term as 
\be
M^{2} = a^{2} m^{2} - \dfrac{z''}{z}, 
\ee where $m^2=3 H^2 \eta_{\mathcal{R}}$, with $\eta_{\mathcal{R}}$ given by \eqref{mass_term}, and $z''/z$ is the usual effective mass term coming from the interaction between $\mathcal{R}$ and the cosmological background.

Varying the action \eqref{SOA_v} and using the Fourier expansion
\begin{equation}
v( \tau, \textbf{x} ) = \int \dfrac{\text{d}^{3} k}{(2 \pi)^{3}} v_{\textbf{k}} (\tau) e^{i \textbf{k}. \textbf{x}},
\end{equation} it is straightforward to obtain

\begin{equation}
v''_{k} + \left( k^{2} + M^{2}\right) v_{k} = 0.
\label{MukhSassa_Eq}
\end{equation} 
Furthermore, this equation can be arranged in the way 
\be
v''_{k} + \left[ k^{2} -\frac{1}{\tau^2}\left(\tilde{\nu}^2-\frac{1}{4}\right)\right] v_{k} = 0,
\label{MukhSassa_Eq2}
\ee where we have defined 
\be
\tilde{\nu}=\nu-\eta_{\mathcal{R}}=\frac{3}{2}+\epsilon+\frac{1}{2}\eta-\eta_{\mathcal{R}}.
\ee For $\tilde{\nu}$ constant and real, the exact solution to \eqref{MukhSassa_Eq2} is 
\be
v_{k}(\tau)=\sqrt{-\tau}\left[C_{1}H_{\tilde{\nu}}^{(1)}(-k\tau)+C_{2}H_{\tilde{\nu}}^{(2)}(-k\tau)\right],
\ee where $H_{\tilde{\nu}}^{(1)}$ and $H_{\tilde{\nu}}^{(2)}$ are the Hankel's functions of first and second kind, respectively \cite{Riotto:2002yw}. By imposing the Bunch-Davies vacuum, such that the solution matches plane-wave solution $v_{k}(\tau)=e^{-ik \tau}/\sqrt{2k}$, at the ultraviolet regime $k\gg a H$ ($-k\tau\ll 1$), and using the relations
\be
\lim_{k\tau\rightarrow -\infty}H_{\tilde{\nu}}^{(1,2)}(-k\tau)=\sqrt{\frac{2}{\pi}}\frac{1}{\sqrt{-k\tau}}e^{\mp i k\tau}e^{\mp i \frac{\pi}{2}\left(\tilde{\nu}+\frac{1}{2}\right)},
\ee we find $c_{1}=\frac{\sqrt{\pi}}{2} e^{i \frac{\pi}{2}\left(\tilde{\nu}+\frac{1}{2}\right)}$ and  $c_{2}=0$.
Therefore, the exact solution to \eqref{MukhSassa_Eq2} becomes
\be
v_{k}(\tau)=\frac{\sqrt{\pi}}{2}e^{i \frac{\pi}{2}\left(\tilde{\nu}+\frac{1}{2}\right)}(-\tau)^{\frac{1}{2}}H_{\tilde{\nu}}^{(1)}(-k\tau).
\ee 
On super-horizon scales $k\ll a H$ ($-k\tau\rightarrow 0$), and using 
\be
\lim_{-k\tau\rightarrow 0} H_{\tilde{\nu}}^{(1)}(-k\tau)=\sqrt{\frac{2}{\pi}}e^{-i\frac{\pi}{2}} 2^{\tilde{\nu}-\frac{3}{2}}\frac{\Gamma(\tilde{\nu})}{\Gamma(\frac{3}{2})} \left(-k\tau\right)^{-\tilde{\nu}},
\ee one finds
\be
v_{k}(\tau)=e^{i \frac{\pi}{2}\left(\tilde{\nu}-\frac{1}{2}\right)}2^{\tilde{\nu}-\frac{3}{2}}\frac{\Gamma(\tilde{\nu})}{\Gamma(\frac{3}{2})}\frac{1}{\sqrt{2 k}} \left(-k\tau\right)^{\frac{1}{2}-\tilde{\nu}}.
\ee Now, taking into account that $\tau=(-1/aH)(1+\epsilon)$ (at first-order) and $\mathcal{R}_{k}=z^{-1} v_{k}=(H/k)(k/aH)(2 Q_{s})^{-1/2} v_{k}$, we write
\bea
\left|\mathcal{R}_{k}\right| &\simeq & \frac{H}{2 \sqrt{ k^3 Q_{s}}}\left(\frac{k}{a H}\right)^{\frac{3}{2}-\tilde{\nu}},\nonumber\\
&\simeq &\frac{H_{k}}{ 2 \sqrt{ k^3 Q_{sk}}}\left[1+\eta_{\mathcal{R}}\ln\left(\frac{k}{a H}\right)\right],
\eea where $H_{k}$ and $Q_{sk}$ are the values of $H$ and $Q_{s}$ at $k=a H$.

The scalar power spectrum of curvature perturbation is calculated as
\bea
\mathcal{P}_{s}(k)&\equiv &\frac{k^3}{2 \pi^2}\left|\mathcal{R}_{k}(\tau)\right|^2,\nonumber\\
&\simeq &\frac{H_{k}^2}{8 \pi^2 Q_{sk}}\left[1+2\eta_{\mathcal{R}}\ln\left(\frac{k}{a H}\right)\right].
\eea Given that $\eta_{\mathcal{R}}\sim \mathcal{O}(\epsilon)$, the consequence of local Lorentz violation is a slight logarithmic time-dependence of the curvature perturbation and its power spectrum at superhorizon scales. Thus, as a satisfactory approximation, we can evaluate this latter at the horizon crossing \cite{Riotto:2002yw}.  

Finally, the scale-dependence of the scalar power spectrum is
\be
n_{s}-1\equiv \left.\frac{d \ln{\mathcal{P}_{s}(k)}}{d\ln{k}}\right|_{k=a H}=-2\epsilon-\eta+2 \eta_{\mathcal{R}}.
\label{ns_fTphi}
\ee This carries out the effects of local Lorentz violation on the scalar power spectrum through the term $2 \eta_{\mathcal{R}}$, at first-order in slow-roll approximation.

\section{Tensor perturbations}\label{tensor_pert}

From the ADM decomposition for the tetrad field presented in Eqs. \eqref{ADM1} and \eqref{ADM2}, 
and using the uniform field gauge, $\delta \phi=0$, we take \cite{Wu:2011kh}
\be
N=1,\:\:\: N^{a}=0, \:\:\: h^{a}_{~i}=a(\delta^{a}_{~i}+\frac{1}{2}\gamma^{a}_{~i}).
\ee 
Then the induced $3-$metric is 
\bea
g_{ij}=\eta_{ab} h^{a}_{~i} h^{b}_{~j}=a^2 \left[\delta_{i j}+h_{i j}+\frac{1}{4} \gamma_{k i} \gamma^{k}_{~j}\right],
\eea where we have defined 
\be
h_{i j}=\frac{1}{2}\eta_{ab}\left(\delta^{a}_{~i}\gamma^{b}_{~j}+\delta^{b}_{~j}\gamma^{a}_{~j}\right)=\frac{1}{2}\left(\gamma_{i j}+\gamma_{j i}\right), 
\ee and $\gamma^{a}_{~j}=\gamma^{i}_{~j}\delta^{a}_{~i}$. Given that the $\gamma^2$ term has contribution only in cubic calculations of the Lagrangian \cite{Maldacena:2002vr}, we keep only until the second term $h_{i j}$ in the induced metric. Also, the tensor $\gamma_{i j}$ can be splitted in the form $\gamma_{i j}=\gamma_{\left(i,j\right)}+\gamma_{\left[i,j\right]}$. The symmetric part $h_{ij}=\gamma_{\left(i,j\right)}$ fulfills the transverse and traceless conditions, $\partial^{i}h_{ij}=h^{i}_{i}=0$, to be gauge invariant \cite{MukhanovBook}. On the other hand, the antisymmetric part matches the gauge degrees of freedom in the local Lorentz invariant theory, and then we identify $B_{ij}$ with $\gamma_{\left[i,j\right]}$.

Then, using the tetrad formalism we find the second-order action for the tensor modes, $h_{ij}=h_{+}e^{+}_{ij}+ h_{\times}e^{\times}_{ij}$, in the way 
\be
S_{T}=\sum_{\lambda} \int{dt d^3x a^3 Q_{T}\left[\dot{h}_{\lambda}^2-\frac{c_{T}^2}{a^2} \left(\partial h_{\lambda}\right)^2\right]},
\label{Tensor_Modes}
\ee where two polarization states are given by $\lambda=+,\times$.  We have also defined
\be
Q_{T}= - \dfrac{1}{2} f_{,T},
\ee
and the squared tensor propagation speed is
\be
c_{T}^2= 1.
\ee The non-ghost condition is satisfied only for $f_{,T}<0$. Besides the usual transverse massless graviton modes, propagating at speed of light, there are no additional propagating modes in the quadratic action \eqref{Tensor_Modes}, which is consistent with local Lorentz invariance of tensor perturbations \cite{Gonzalez-Espinoza:2019ajd}. 

The power spectrum for tensor perturbations becomes
\be
\mathcal{P}_{T}=\frac{H_{k}^2}{2 \pi^2 Q_{Tk}},
\ee with $H_{k}$ and $Q_{Tk}$ the values of $H$ and $Q_{T}$ at $k=a H$.
Thus, the spectral index is 
\be
n_{T}\equiv \left.\frac{d \ln{\mathcal{P}_{T}}}{d \ln{k}}\right|_{k=aH}=-2\epsilon-\delta_{f_{,T}}.
\label{nT}
\ee 

Tensor-to-scalar ratio, evaluated at the horizon crossing, is given by

\begin{equation}
r = \dfrac{\mathcal{P}_{T}}{\mathcal{P}_{s}} \simeq 16 \delta_{PX}=16 \left(\epsilon-\delta_{f_{,T}}\right).
\label{r}
\end{equation} Using the Eqs. \eqref{nT} and \eqref{r}, we obtain the consistency relation
\be
r= 8 \left(-n_{T}-3 \delta_{f_{,T}}\right).
\label{r_2}
\ee This is agreement with the standard inflation limit where $r= -8 n_{T}$. The quantity $\delta_{f_{,T}}$ appears as a small correction to the value of standard inflation. According with Eq. \eqref{slow_roll_eps}, and since $\epsilon$ and $\delta_{PX}$ are small during slow-roll inflation, it is expected that the parameter $\delta_{f_{,T}}$ in Eq. \eqref{deltafT} is also small. This condition must be satisfied for function $f(T,\phi)$ in order to support the slow-roll inflationary scenario. However, although $\delta_{f_{,T}}$ is of first order in slow-roll approximation, it can cause a significant change to $r$ in Eq. \eqref{r_2}, which therefore can also be contrasted with observational data.

\section{Application to a concrete model of inflation}\label{Chaotic}

We consider the ansatz
\be
f(T,\phi)=-\frac{M_{pl}^2}{2} T-G(T) F(\phi)-V(\phi),
\label{Part_Model}
\ee with  $G(T)=T^s$, $F(\phi)=\xi \phi^c$, and $V(\phi)=\lambda \phi^d$, where $s$, $c$, $d$, $\xi$ and $\lambda$ are positive constants.

Under the slow-roll approximation, $\dot{\phi}^2/2\ll V$ and $\abs{\ddot{\phi}}\ll H \abs{\dot{\phi}}$ \cite{MukhanovBook}, the backgrounds equations \eqref{00} and \eqref{phi} give
\be
 T\simeq \left[\frac{\lambda}{\xi\left(2s-1\right)}\right]^{\frac{1}{s}}\phi^{\frac{d-c}{s}},
\label{T_High_E_L}
\ee with $s\neq 1/2, 0$, and $\phi$ becomes  
\be
\frac{\phi^{2+\frac{d-c}{s}-d}}{2+\frac{d-c}{s}-d}\simeq 2\lambda \left[\frac{\xi\left(2s-1\right)}{\lambda}\right]^{\frac{1}{s}}\left(\frac{c}{2 s-1}+d\right) N,
\label{phi(N)}
\ee where we have introduced the e-folds number $N$ \cite{MukhanovBook}. Here, we have also applied the high energy limit $F G/(M_{pl}^2 T)\gg 1$, for $N\gg 1$, with  $\mu\simeq TG_{,TT}/G_{,T}= s-1$. Thus, from Eq. \eqref{mass_term}, we obtain
\bea
 \eta_{\mathcal{R}}&=&-\frac{2 \lambda}{s}\left[\frac{\xi\left(2 s-1\right)}{\lambda}\right]^{\frac{1}{s}} \left(\frac{c}{s-1}+d\right)\times \nonumber\\
&& \left[\frac{c (3 s-2)}{2 s-1}+2 d (s-1)\right] \phi^{\frac{c-d}{s}+d-2},
\label{mass_term_High_E}
\eea which is divergent for $s=1$. 

In the high energy (high curvature) regime for $N\gg1$, and near the horizon crossing, the torsion scalar $T$ takes a larger value compared with its value in the end of inflation, and then it is expected that higher powers than the linear case on $T$ are preferred when the modified gravity terms are dominant over the Einstein-Hilbert one. In this way, as we are interested in inflation driven by these modified gravity corrections in this high energy limit, this brings us to assume $s>1$  \cite{Clifton:2011jh}. Ultimately, a power less than $s=1$ ($s\neq 1/2$ in Eq. \eqref{T_High_E_L}) could also be considered with the risk that modified gravity terms become dominant at the end of inflation and then possibly spoiling reheating after inflation \cite{DeFelice:2010aj}. Furthermore, since we have applied this high energy approach from equation \eqref{T_High_E_L}, the GR limit with minimally coupled scalar field, for $s=0$, has also been excluded from the present analysis in equation \eqref{mass_term_High_E}. We can also highlight that in order to study slow-roll inflation in model \eqref{Part_Model}, without applying this high energy limit, would require a complete numerical integration of the slow-roll equations\cite{Gonzalez-Espinoza:2019ajd}. This latter study lies beyond the scope of the present work.

The scalar power spectrum is written as
\be
\mathcal{P}_s= \frac{(2 s-1)^{2-\frac{3}{s}} \lambda ^{\frac{3}{s}-2} \phi^{\frac{3 (d-c)}{s}-2 d+2}}{96 \pi ^2 \xi ^{\frac{3}{s}} (c+d (2 s-1))^2},
\ee 
whereas $n_{s}$ and $r$ are given by
\bea
n_{s} &=& 1-\frac{2\lambda}{(s-1) s} \left[\frac{\xi\left(2 s-1\right)}{\lambda}\right]^{\frac{1}{s}}\times \nonumber\\
&& \Bigg[\frac{c^2 (3 s-1)}{2 s-1}+\frac{2 c (s-1) (3 d s-d+s)}{2 s-1}+\nonumber\\
&& d (s-1)(2 (d+1) s-d)\Bigg] \phi^{\frac{c-d}{s}+d-2},\\
r &=& \frac{16\lambda\left(2 s-1\right)}{s} \left[\frac{\xi\left(2 s-1\right)}{\lambda}\right]^{\frac{1}{s}} \times \nonumber\\
&& \left(\frac{c}{2s-1}+d \right)^2  \phi^{\frac{c-d}{s}+d-2}.
\eea  
Using Eq. \eqref{phi(N)}, and for $N\gg 1$, $n_{s}$ and $r$ take the form $n_{s}=1-p/N$ and $r=q/N$ with $p=p(s,c,d)$ and $q=q(s,c,d)$, functions of $s$, $c$ and $d$. The latest cosmological data from Planck satellite \cite{Akrami:2018odb} fixed the values of $n_{s}$ and $r$ in the ranges $n_{s}=0.9649\pm 0.0042$ at $68\%$ CL, and $r<0.064$, at $95\%$ CL, which allows us constrain the parameters of $s$, $c$ and $d$. Additionally, from the observational data for the scalar power spectrum, $\mathcal{P}_{s}=2.141\times 10^{-9}$ \cite{Akrami:2018odb}, we can obtain an estimate for $\lambda$ and $\xi$. For $c=0$, which corresponds to $f(T)$ gravity plus scalar field, and for $N=50$, it is required $s>2.318$, and $\frac{0.3075 s}{s-0.71784}\leq d<\frac{4 s}{12 s-7}$. For example, for $s=3$, we have $0.4042\leq d<0.4138$, and then $-0.0064<\eta_{\mathcal{R}}\leq -0.0062$, with the minimum value of $r$ as $r\simeq 0.062$. Smaller values for $r$ can be obtained in the range $r\in \left[0.058,0.059\right]$, for $s\gtrsim 12$ and $-0.0080\lesssim \eta_{\mathcal{R}}\lesssim -0.0071$, with $n_{s}$ still inside the observational bounds. We also obtain $\lambda/M_{pl}^{4-d} \sim 10^{-7}$ and $\xi M_{pl}^{2} \gtrsim 10^{15}$. These results are consistent with the observational data found for the concave potential, $V_{,\phi\phi}<0$, in Ref. \cite{Akrami:2018odb}. However, in this case, since $d<1$, the stability condition of de-Sitter background in Eq. \eqref{deSitterStability} is not satisfied. 

\begin{figure}[h]
	\centering
		\includegraphics[width=0.4\textwidth]{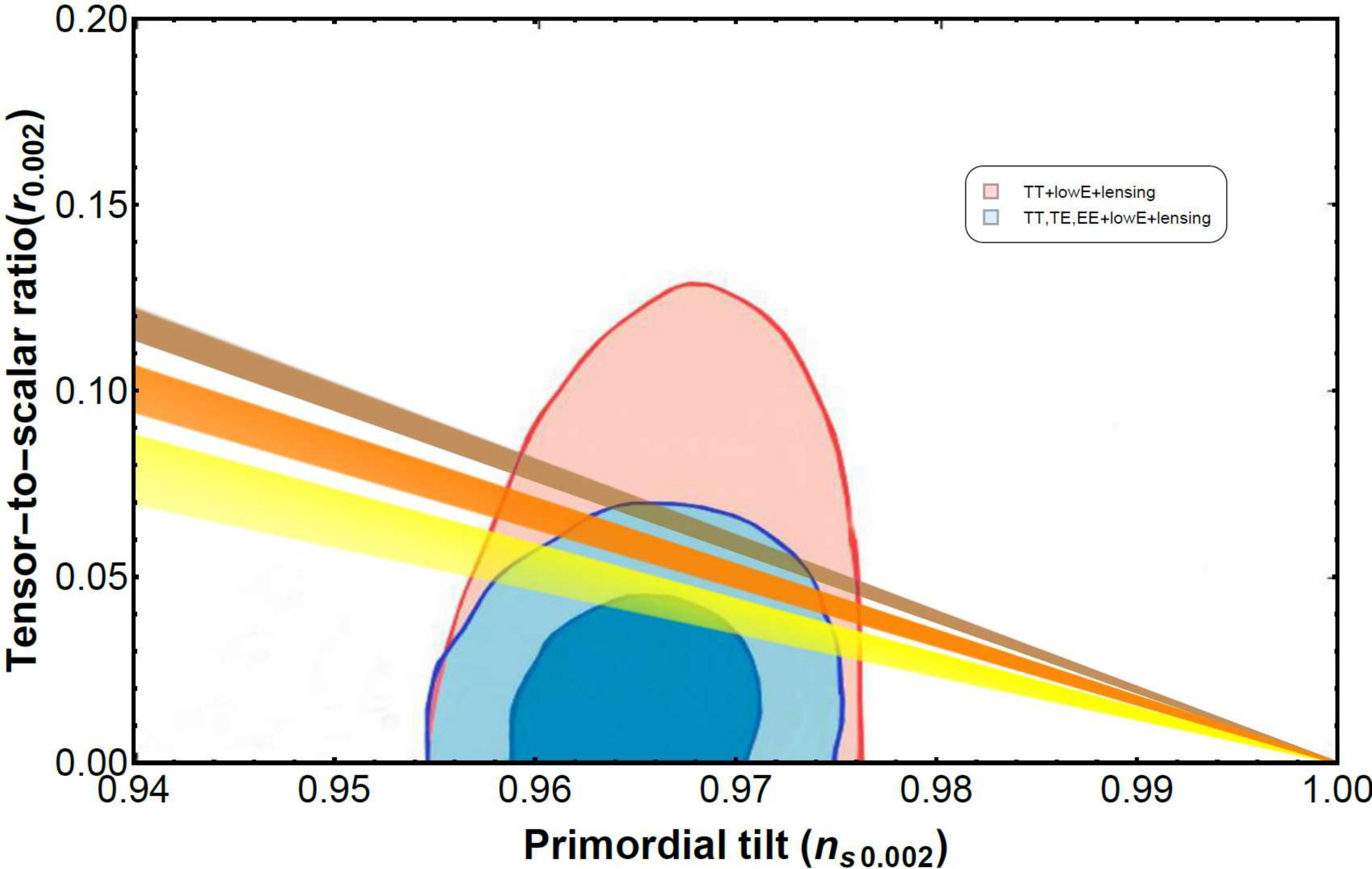}
	\caption{It is shown the region swept by the curve $r(n_{s})$ for $f(T,\phi)=-M_{pl}^2 T/2-\xi T^s\phi^c-\lambda \phi^d$, consistent with the two-marginalized constraints jointly as $68\%$ and $95\%$ C.L. at $k=0.002~Mpc^{-1}$, from the Planck 2018 results \cite{Akrami:2018odb}. The brown region corresponds to $c=0$, $s \in \left[3,20\right]$ and $d\in \left[0.3189, 0.4138\right]$, while the yellow region corresponds to $c=0.01$, $d=2/3$, and $s\in \left[1.0002, 1.0003\right]$, and the orange region is associated with $c=0.05$, $d=1$, and $s\in \left[1.002,1.003\right]$.}
	\label{FIG1}
\end{figure}

\begin{figure}[h]
	\centering
		\includegraphics[width=0.4\textwidth]{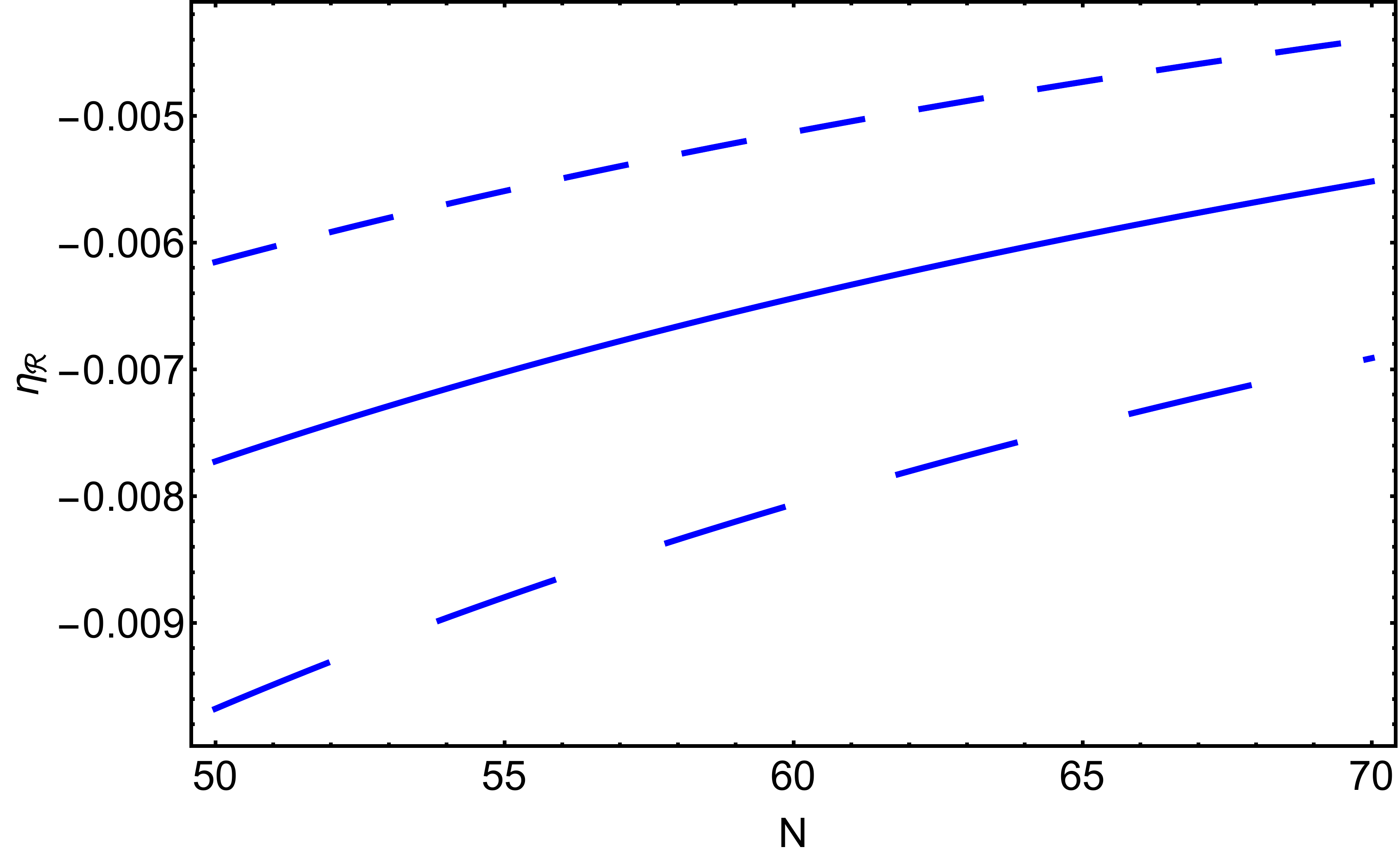}
	\caption{It is depicted the behaviour of $\eta_{\mathcal{R}}=m^2/(3 H^2)$ as function of $N$, for model $f(T,\phi)=-M_{pl}^2 T/2-\xi T^s\phi^c-\lambda \phi^d$. The short-dashed line corresponds to $s=3$, $c=0$ and $d=0.4$, while the solid one is related to $s=1.0002$, $c=0.01$, and $d=2/3$, and the long-dashed line happens for $s=1.003$, $c=0.05$, and $d=1$. }
	\label{FIG2}
\end{figure}

For non-minimally coupled scalar-torsion models, with $c>0$, it is required $c<d$, in order to have $\dot{H}<0$. Using the observational data of $n_{s}$ and $r$, and for $N=70$, and $d=1$, with $c=0.05$,  we find the range $1.00214\leq s\leq 1.00445$, and $-0.00924\leq \eta_{\mathcal{R}}\leq -0.00504$. On the other hand, for $N=60$, and $d=2/3$, with $c=0.01$, we get $1.00015\leq s\leq 1.0003$ and $-0.00850\leq \eta_{\mathcal{R}} \leq -0.00438$. In this latter case we obtain the minimum value of $r$ as $r\simeq 0.045$, with $n_{s}$ still inside the observational bounds, which is compatible with the current Planck data at $68\%$ CL. For these values we get $\lambda/M_{pl}^{4-d} \sim 10^{-8}$ and $\xi M_{pl}^{2+c} \sim 10$.  In FIG. \ref{FIG1}, we show the curves $r(n_{s})$ for several different values of the parameters. In FIG. \ref{FIG2}, it is depicted the behaviour of $\eta_{\mathcal{R}}$ as function of $N$. It can be verified that $\abs{\eta_{\mathcal{R}}}=\abs{m^2}/(3 H^2)\ll 1$.

%%%%%%%%%%%%%%%%%%%%%%%%%%%%%%%%%%%%%%%%%%%%%%%%%%%%%%%%%%%%
\section{Conclusions}\label{Concluding_Remarks}

We have studied the generation of primordial fluctuations in generalized teleparallel scalar-torsion gravity theories whose  Lagrangian density is an arbitrary function $f(T,\phi)$ of the torsion scalar $T$ and a scalar field $\phi$, plus the kinetic term of this latter. To develop primordial density perturbations, we started from the Arnowitt-Deser-Misner (ADM) formalism of the tetrad field, and we choose the uniform gauge \cite{Baumann:2014nda}. The tetrad field has sixteen degrees of freedom and local Lorentz invariance of TG allows us to eliminate six degrees of freedom, yielding the same number of independent components of the metric tensor \cite{Aldrovandi-Pereira-book}. However, it is well known that the action for MTG is no longer a local Lorentz invariant, and thus, the field equations are not completely symmetric \cite{Sotiriou:2010mv,Li:2010cg}. In order to restore the local Lorentz invariance, we have introduced six additional degrees of freedom in the form of Goldstone modes of the symmetry breaking, through a Lorentz rotation of the tetrad field \cite{Bluhm:2007bd}. So, the antisymmetric part of field equations constitutes a set of six equations for six extra modes, that is, a scalar, a transverse $3$-vector, and a spatial antisymmetric tensor modes. 

Putting all these pieces together, and after integrating out the auxiliary fields, we have calculated the second order action for the propagating modes \cite{Maldacena:2002vr}. As usual, we have treated the scalar and tensor modes separately since they are not coupled. Vector modes decay rapidly with the cosmic expansion, and thus they can be ignored \cite{MukhanovBook}. Furthermore, we have verified that the corresponding additional tensor modes are completely cancelled out from the second order action for tensor perturbations, remaining only the usual transverse massless graviton modes, propagating at speed of light, and therefore indicating the local Lorentz invariance in the tensor perturbations sector \cite{Gonzalez-Espinoza:2019ajd}.

In the second order action for the curvature perturbation \eqref{SecOrderSM}, it is observed the emergence of an explicit mass term, $\eta_{\mathcal{R}}=m^2/(3H^2)$, which represent the effects of local Lorentz violation. This explicit mass term is of first-order in slow-roll approximation, and it is always nonzero (and finite), for nonminimal coupling functions $f(T,\phi)$, which are non-linear in $T$, including the $f(T)$ gravity, plus scalar field, as a particular example.  The arising of this propagating massive scalar mode could be related to an alternative Higgs mechanism that has no direct analogue in nonabelian gauge theory \cite{Kostelecky:1989jw}. As expected, in the case of TG we obtain $\eta_{\mathcal{R}}=0$, because the local Lorentz invariance of the theory \cite{Aldrovandi-Pereira-book}. On the other hand, when the nonminimal coupling function $f(T,\phi)$ is linear in $T$, like the action considered in \cite{Wu:2016dkt}, it becomes divergent, which necessarily leads us to $\partial^{2}{\mathcal{R}}=0$, or equivalently, $\mathcal{R}_{k}=0$ for all Fourier mode $k$, and hence, it is immediate to conclude that no subhorizon scalar mode could propagate and survive by the time of horizon crossing. This latter result is consistent with what was obtained in \cite{Wu:2016dkt}.

Our results indicate that only for MTG theories with non-linear coupling functions $f(T,\phi)$, including the $f(T)$ gravity, plus scalar field, there will be generation of primordial density fluctuations to be contrasted with observations. At sub-horizon scales, the solution for the primordial quantum fluctuations matches the Bunch-Davies vacuum boundary condition, and  then at scales deep inside the horizon, the effects of $\eta_{\mathcal{R}}$ can be neglected. At super-horizon scales, the curvature perturbation modes freeze up to a slight logarithmic time-dependence proportional to slow-roll parameters, and thus the spectral index of scalar power spectrum evaluated at the horizon crossing, Eq. \eqref{ns_fTphi}, carries out the effects of local Lorentz violation. For large field inflationary models, and using the non-ghost condition $f_{,T}<0$, the explicit mass term could contribute with the red tilt of the scalar spectrum, but tachyonic instability is avoided as long as the instability rate is less than the Hubble rate \cite{DeFelice:2016ucp,Frusciante:2018vht}. We have applied our results to chaotic inflation and corroborated them by using the current Planck data \cite{Akrami:2018odb}.

Finally, we note that the slow-roll parameter $\delta_{f_{,T}}=\dot{f}_{,T}/(H f_{,T})$ can cause a significant change to the tensor-to-scalar ratio $r$ in Eq. \eqref{r_2}. Particularly, the contribution of $\delta_{f_{,T}}$ to $r$ can either lower its value bringing it to values more compatible with observations, or raise it too much and then leaving it outside the allowed contour regions from the latest Planck data. So, the effect of $\delta_{f_{,T}}$ on $r$ and the contribution of $\eta_{\mathcal{R}}$ to the spectral index $n_{s}$ in Eq. \eqref{ns_fTphi}, are interesting results which could indicate the safe signature of new physics, and then providing us a smoking gun for these kinds of teleparallel modifications when comparing their theoretical predictions with the observational data.
 
%%%%%%%%%%%%%%%%%%%%%%%%%%%%%%%%%%%%%%%%%%%%%%%

%\balance
%\clearpage

%%%%%%%%%%%%%%%%%%%%%%%%%%%%%%%%%%%%%%%%%%%%%%%%%%%%%%%%%%%%%%%%%%%%
\section*{Acknowledgements}
The authors would like to thank Joel Saavedra and Nelson Videla for invaluable discussions and suggestions. M. Gonzalez-Espinoza acknowledges support from PUCV. G. Otalora acknowldeges DI-VRIEA for financial support through Proyecto Postdoctorado $2020$ VRIEA-PUCV.
%%%%%%%%%%%%%%%%%%%%%%%%%%%%%%%%%%%%%%%%%%%%%%%%%%%%%%%%%%%%%%%%%%%%

%\clearpage
%\appendix


%\clearpage

\bibliography{bio}   % name your BibTeX data base

\end{document}